\documentclass[journal=jacsat,manuscript=article]{achemso}

\usepackage[version=3]{mhchem} 
\usepackage{color}
\usepackage{braket}

\usepackage{xcolor}
\usepackage{changes}

\definecolor{acolour}{RGB}{155, 25, 55}

\definechangesauthor[name={Aleksei}, color={acolour}]{AVI}


\author{Aleksei V. Ivanov}
\affiliation{Science Institute and Faculty of Physical Sciences, University of Iceland VR-III, 107 Reykjav\'{\i}k, Iceland.}
\alsoaffiliation{St. Petersburg State University, 199034, St. Petersburg, Russia}
\author{Tushar K. Ghosh}
\affiliation{Science Institute and Faculty of Physical Sciences, University of Iceland VR-III, 107 Reykjav\'{\i}k, Iceland.}
\alsoaffiliation{Department of Applied Physics, Aalto University, Espoo, FI-00076, Finland.}
\author{Elvar \"O. J\'onsson}
\affiliation{Science Institute and Faculty of Physical Sciences, University of Iceland VR-III, 107 Reykjav\'{\i}k, Iceland.}
\alsoaffiliation{Department of Applied Physics, Aalto University, Espoo, FI-00076, Finland.}
\author{Hannes J\'onsson}
\email{hj@hi.is}
\affiliation{Science Institute and Faculty of Physical Sciences, University of Iceland VR-III, 107 Reykjav\'{\i}k, Iceland.}
\alsoaffiliation{Department of Applied Physics, Aalto University, Espoo, FI-00076, Finland.}
 
\title{Mn Dimer can be Described Accurately with Density Functional Calculations when Self-interaction Correction is Applied}

\begin{document}


\begin{tocentry}
   \includegraphics[width = 0.45\columnwidth]{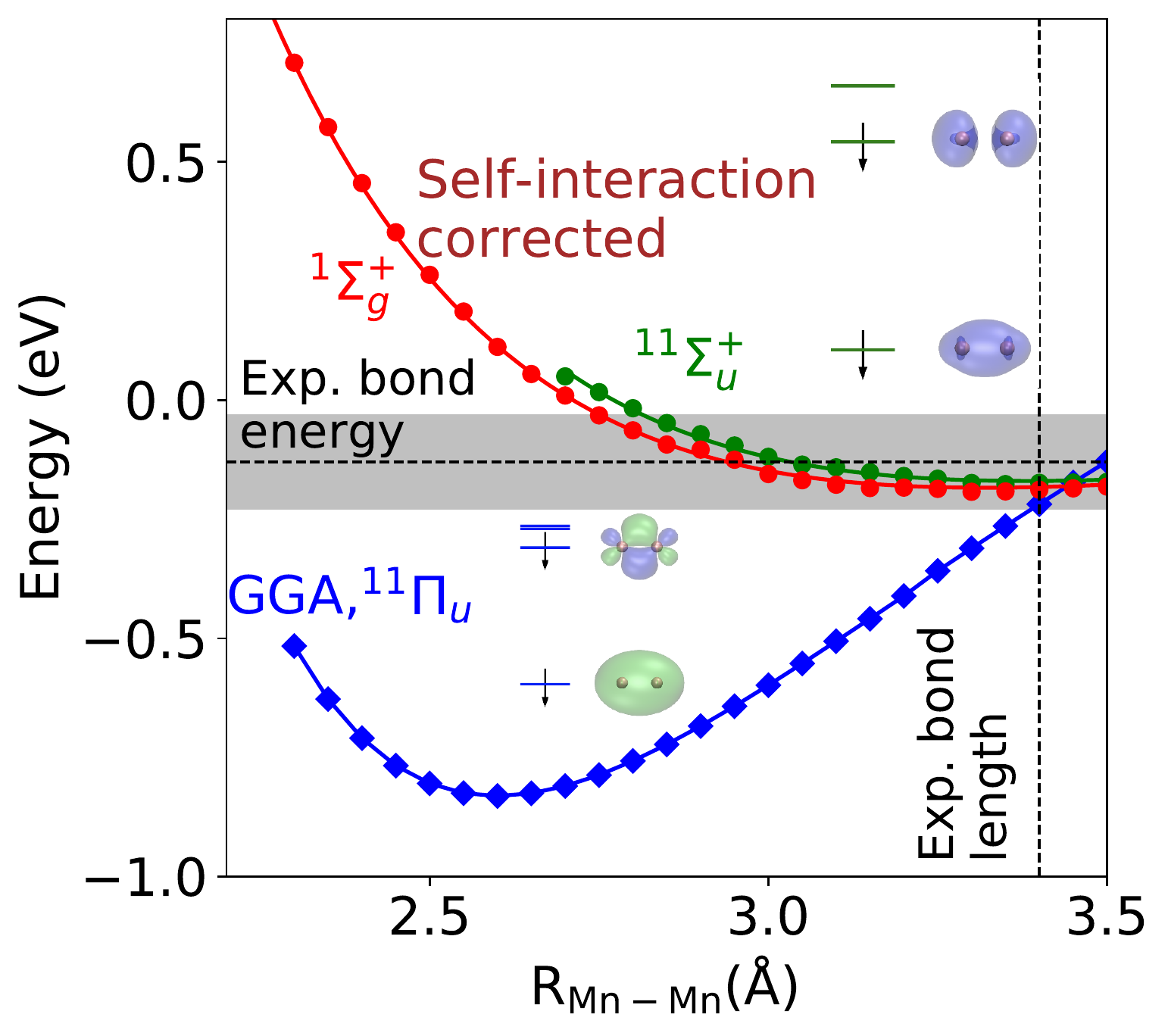}
\end{tocentry}

\newpage
\begin{abstract}
Qualitatively incorrect results are obtained for the Mn dimer in density functional theory calculations
using the generalized gradient approximation (GGA) and similar results are obtained from local density and meta-GGA functionals.
The  coupling is predicted to be ferromagnetic rather than antiferromagnetic and the 
bond between the atoms is predicted to be an order of magnitude too strong and about an {\AA}ngstrøm too short. 
Explicit, self-interaction correction (SIC) applied to a commonly used GGA energy functional, however, provides close agreement with 
both experimental data and high-level, multi-reference wave function calculations.
These results show that the failure is not due to strong correlation but rather the single electron self-interaction
that is necessarily introduced in estimates of the classical Coulomb and exchange-correlation energy when only the total electron density is used as input.
The corrected functional depends explicitly on the orbital densities and can, therefore, avoid the introduction of self-Coulomb interaction.
The error arises because of over-stabilization of bonding $d$-states in the minority spin channel 
resulting from an overestimate of the $d$-electron self-interaction in the
semi-local 
exchange-correlation functionals.
Since the computational effort in the self-interaction corrected calculations scales with system size in the same way as for regular semi-local functional calculations, 
this approach provides a way to calculate properties of Mn nanoclusters as well as biomolecules and extended solids where Mn dimers and larger cluster are present,
while multi-reference wave function calculations can only be applied to small systems.  
 \end{abstract}

\newpage
Manganese atoms have large magnetic moment and are, therefore, of interest for various technical applications. 
Complexes of Mn atoms are also found in several metalloenzymes\cite{Book_EPR_bio,Dismukes_chem_rev}, 
for example in the oxygen-evolving photo-system II.  
The properties of such systems are of great interest and theoretical calculations could in principle provide valuable information to help gain
an understanding of the role Mn atoms play.
However, theoretical calculations prove to be particularly challenging for these systems. 
%
%
%
%
%
The Mn dimer is the simplest manganese complex 
and it represents an important test system for theoretical methods that could ultimately be used for the larger and more complex systems.
Its properties are quite well known from electron spin resonance and optical absorption measurements of dimers confined in a rare gas matrix.
The ground state is found to be antiferromagnetic\cite{rivoal1982,baumann1983,kirkwood1991}
with a small bond energy of 0.13$\pm0.1$ eV\cite{kant1968}
and a large bond length of 3.4 {\AA}.\cite{baumann1983,cheeseman1990}
Resonance Raman spectra give vibrational frequency of 76 cm$^{-1}$ in a Kr matrix\cite{bier1988} 
and 68 cm$^{-1}$ in a Xe matrix.\cite{kirkwood1991}

High-level wave function calculations of an isolated Mn dimer give results in close agreement with the experimental measurements.
Both complete-active-space self-consistent field 
in combination with second-order
perturbation theory\cite{yamamoto2006}, as well as multi-reference 
\cite{wang2004,yamamoto2006,camacho2008,angeli2008,buchachenko2010} calculations have been carried out.
They predict bond energy in the range 0.10 - 0.14 eV and bond length of 3.3 - 3.8 {\AA} in an antiferromagnetic ground state with a coupling constant of 
J=-5.8 cm$^{-1}$.
It is clear from the close agreement between these calculations and the experimental measurements that the effect of the confining rare gas matrix is small and
the measurements indeed probe the properties of the Mn dimer.
Such high-level, wave function based calculations become, however, impractical for larger systems due to the strong scaling of the 
computational effort with system size.

Kohn-Sham density functional theory (KS-DFT)\cite{kohn1965}
can provide a valuable tool for theoretical studies of large systems with up to several hundred atoms, free -- in principle -- of adjustable parameters with unknown values.
Unfortunately, the results of such calculations for the Mn dimer with commonly used energy functionals such as 
local density approximation (LDA) and generalized gradient approximation (GGA), 
are in strong disagreement with the experimental measurements and the high-level, wave function calculations.\cite{pederson1998,Yanagisawa2000,barborini2016}
%
%
The ground state is predicted to be ferromagnetic rather than antiferromagnetic
with the bond between the Mn atoms being much too strong and too short.
All electron calculations with various GGA functionals give a binding energy of around 0.9 eV and bond length of 2.57-2.61 {\AA}.\cite{barborini2016}
Below, we present results using an elaborate meta-GGA functional, SCAN,\cite{sun2015}
where the Mn dimer is also found to be poorly described.
We will from now on refer to calculations using LDA, GGA or meta-GGA functionals collectively as semi-local DFT calculations.

Similar failure in DFT calculations of magnetic coupling constants has also been reported for various manganese binuclear
complexes.\cite{Pantazis_Inorganics}
Shortcomings of DFT calculations are often ascribed to `strong correlation' and systems where large errors are obtained 
are often characterized as `highly correlated systems' (for a recent discussion of a possible meaning of these terms, see Ref. \cite{perdew2021}).
The Mn dimer is an example of such a system.
It has, however, been shown that calculated results for the Mn dimer can be improved when exact exchange
is added to a semi-local DFT functional to form a hybrid functional, but the weight of the exact exchange in this blend needs to be significantly larger than the
range of 0.20 to 0.25 in commonly used functionals of this form.\cite{yamanaka2007,Pantazis_Inorganics,barborini2016}
Semi-local DFT functionals tend to give errors of opposite sign to those of Hartree-Fock calculations, so some mix of the two can often be tuned to give the desired result.

An alternative reason for the failure of semi-local DFT calculations is the self-interaction error that is necessarily introduced in the estimate of the 
classical Coulomb interaction between the electrons when only the total electron density is used as input. 
This error is highly non-local. In the exchange-correlation part of
the semi-local functionals a correction is estimated, i.e.~a self-interaction contribution of opposite sign, but 
the cancellation can mathematically not be complete and, therefore, self-interaction error remains and can lead to erratic results.
Previously it has been speculated that the self-interaction error is not responsible
for the poor performance of semi-local DFT functionals for the Mn dimer
but rather strong correlation.\cite{pederson1998,Yanagisawa2000}
So, the question remains whether the large error in semi-local DFT calculations of the Mn dimer is due to strong correlation
or self-interaction, or possibly some other source of error.

In this letter, the results of variational, self-consistent calculations are presented where the self-interaction error is removed explicitly
as proposed by Perdew and Zunger\cite{perdew1981}. 
As described below, the results are found to be in remarkably good agreement with the high-level wave function calculations. 
While any form of correlation necessarily reflects interaction between two or more electrons, the self-interaction error 
in semi-local DFT functionals is present even for systems containing a single electron and the correction applied here
involves terms that depend only on 
one-electron densities.
The results presented here therefore demonstrate that the problem in the semi-local DFT
calculations of the Mn dimer is not related to correlation, but rather self-interaction error. 
In addition to providing important insight into the reason for the failure of DFT for the Mn dimer,
this opens up an avenue for accurate calculations of larger systems containing Mn complexes 
because the computational effort of the GGA calculations with self-interaction correction scales with system size in the 
same way as regular GGA calculations.\cite{klupfel2012b}




In Kohn-Sham density functional theory,\cite{kohn1965} 
the energy of an electronic system is estimated from
\begin{equation}
E^\mathrm{KS}[\rho]  = T_s + \int v_\mathrm{ext}({\bf r}) \rho({\bf r}) d{\bf r} + E_\mathrm{C}[\rho]  + E_{xc}[\rho_{\uparrow},\rho_{\downarrow}]
\end{equation}
were, $T_{s}$ is the kinetic energy of an independent electron system described by spin-orbitals $\phi$ and the
electron density
\begin{equation} 
\rho_\sigma({\bf r}) \ = \ \sum_{i_\sigma} \rho_{i_\sigma}({\bf r}) \ = \ \sum_{i_\sigma} |\phi_{i_\sigma}({\bf r})|^2 
\end{equation}
corresponds to the ground state electron density of the interacting electron system for each spin channel, $\sigma= \{\uparrow,\downarrow\}$.
The energy due to the electron-nuclei interaction, described by the external potential $v_\mathrm{ext}$, can be evaluated correctly from the total electron density,
$\rho({\bf r})=\rho_\uparrow({\bf r})+\rho_\downarrow({\bf r})$,
but the estimate of the classical Coulomb interaction between the electrons
%
\begin{equation}
E_\mathrm{C}[\rho] \ = \  \int\int {{\rho({\bf r}) \rho({\bf r^\prime})} \over {| {\bf r} -{\bf r^\prime}|}} d{\bf r} d{\bf r^\prime}
\end{equation}
includes spurious interaction of each electron with itself.
This is most clearly seen for a system with a single electron where non-zero interaction energy is obtained from this estimate.
A more accurate estimate can be obtained by using the spin-orbital densities, $ \rho_i$, where interaction of a spin-orbital with itself is avoided
%
\begin{equation}
E_\mathrm{C}^\mathrm{SIC}[\rho_1, \dots, \rho_N] \ = \ E_\mathrm{C}[\rho] -  \sum_i  \int\int {{\rho_i({\bf r}) \rho_i({\bf r^\prime})} \over {| {\bf r} -{\bf r^\prime}|}} d{\bf r} d{\bf r^\prime} 
= 
 \sum_{i} \sum_{j>i}  \int\int {{\rho_i({\bf r}) \rho_j({\bf r^\prime})} \over {| {\bf r} -{\bf r^\prime}|}} d{\bf r} d{\bf r^\prime}
\end{equation}
%
Here, the summation indices run over both spin channels, $i= \{i_\uparrow,i_\downarrow\}$.
The exchange-correlation energy term in the KS-DFT functional, $E_{xc}$,
attempts to provide such a correction
but, because of the semi-local form, cannot accurately cancel out the non-local self-interaction error in $E_\mathrm{C}[\rho]$.

Perdew and Zunger proposed a procedure where the net self-interaction error is estimated for each spin-orbital and the sum 
subtracted from the Kohn-Sham functional\cite{perdew1981}
\begin{equation}
E^\mathrm{SIC}[\rho_1, \dots, \rho_N] = E^\mathrm{KS}[\rho]  -  \sum_{i} \left( E_\mathrm{C}[\rho_{i}] +  E_{xc}[\rho_{i}, 0] \right) .
\end{equation}
This provides the correction to the classical Coulomb energy as in Eqn.~(4) and also addresses the extent to which the exchange-correlation
functional is able to cancel out the self-interaction by evaluating the net self-interaction for each spin-orbital separately.
For a one electron system, the corrected functional is guaranteed to be self-interaction free, but for many electron systems, this 
correction procedure is approximate.
While this approach was originally proposed for the LDA functional, it can also be applied to GGA functionals but there it has been found to 
give an overcorrection and a scaling 
by 1/2 has been shown to give good results for a wide range of systems and properties such as atomization energy of molecules, band gaps of solids
and the balance between localized and delocalized electronic states. 
\cite{Jonsson2011,klupfel2012,gudmundsdottir2015}. 
We choose here to use the PBE functional,\cite{perdew1996} a GGA functional approximation that is commonly used in calculations of condensed matter
and refer to the corrected functional as PBE-SIC/2.  As discussed below, the scaling of 1/2 is not essential here, similar results are obtained for the Mn dimer
without the scaling.

The corrected functional is not unitary invariant as it depends explicitly on the orbital densities, as indicated in Eqs.~(4-5),
and it turns out that the optimal spin-orbitals that minimize the energy of the system are hybrid orbitals, 
i.e. linear combinations of the canonical orbitals that are eigenfunctions of the Hamiltonian.
The calculations need to make use of complex-valued functions to represent the optimal spin-orbitals.\cite{klupfel2011,Lehtola2013,Lehtola2016}
Here, a  real-space grid has been used\cite{enkovaara2010} combined with 
projector augmented wave (PAW) to represent the effect of frozen core-electrons of the atoms.\cite{blochl1994}
A localized atomic orbital basis set is used including primitive Gaussians from the def2-TZVPD basis set
\cite{Weigend2005,Rappoport2010,Feller1996,Schuchardt2007,Pritchard2019} 
augmented with a single-zeta basis.\cite{rossi2015}
Tests against calculations using full flexibility of the real-space grid 
with mesh size of 0.15 {\AA} give nearly identical results, showing that the atomic basis set is sufficient. 
Starting from localized orbitals,\cite{Jonsson2017}
the energy is variationally minimized using an exponential transform direct optimization method described elsewhere.\cite{DFTmin2021,DFTmin2020}



\begin{figure}[!t]
\begin{center}
\includegraphics[width = 0.9\columnwidth]{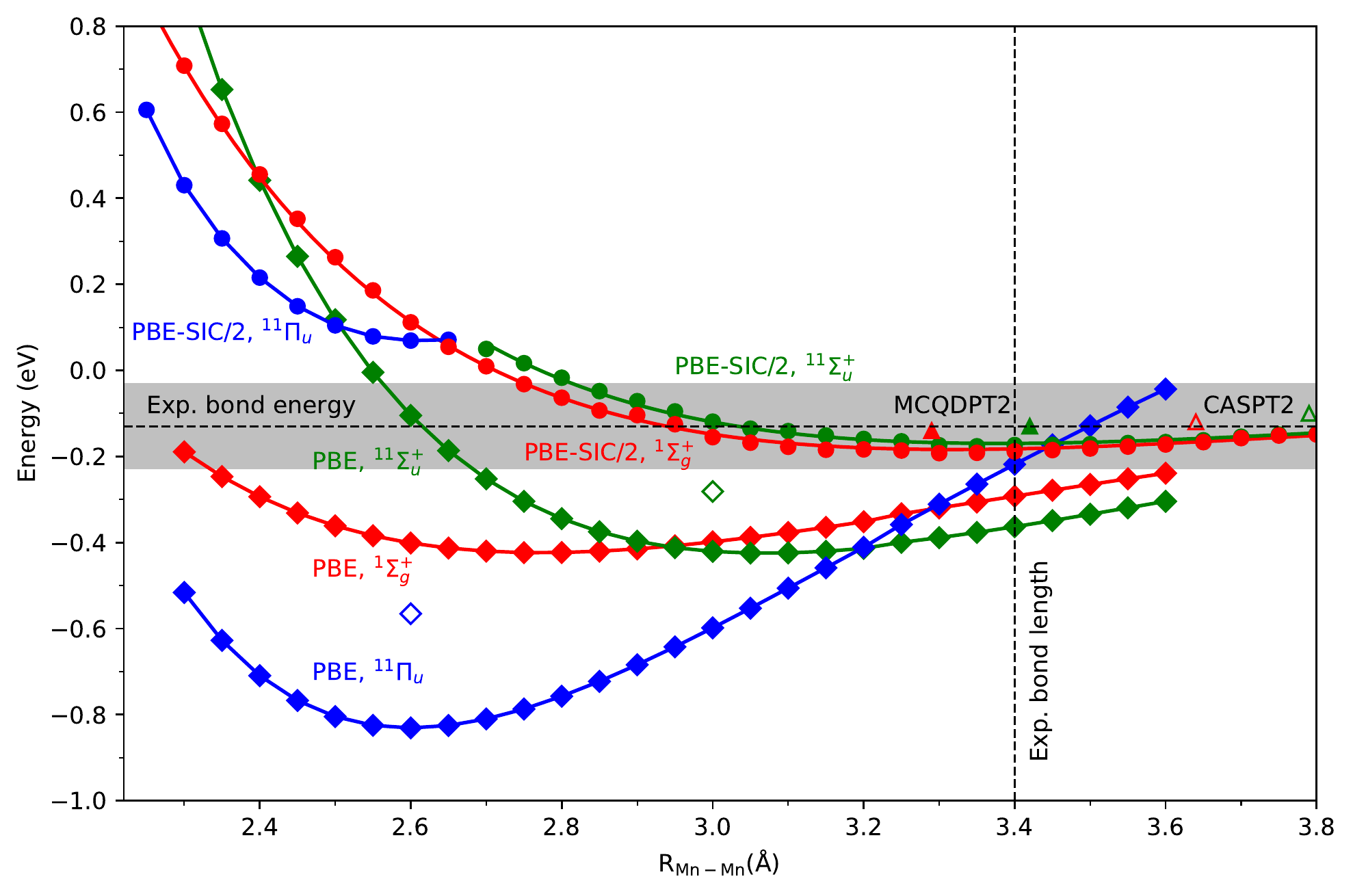}
\caption{
Energy of the Mn dimer in the three lowest lying electronic states as a function of distance between the atoms calculated with the PBE functional (filled diamonds)
and self-interaction corrected PBE-SIC/2 (filled circles). The zero of energy is twice the energy of an Mn atom in the ${}^{6}\mathrm{S}$ electron configuration.
The ferromagnetic states ${}^{11}\Sigma^{+}_{u}$  and ${}^{11}\Pi_{u}$ are shown in blue and green, respectively,
and the antiferromagnetic state ${}^{1}\Sigma_g^+$ in red.
The ground state in the PBE calculations is ferromagnetic with a binding energy of nearly 0.9 eV and bond length of 2.6 {\AA}
while the antiferromagnetic state is the ground state in the self-interaction corrected PBE-SIC/2 calculations, with
binding energy of 0.18 eV and bond length of 3.32 {\AA}.
Experimental estimates\cite{kant1968,baumann1983,cheeseman1990} 
of the bond length and bond energy (shaded grey area indicating the estimated error bar) are shown with dashed lines.
The triangles show results of high-level calculations using MCQDPT2\cite{yamamoto2006} (filled) and CASPT2\cite{wang2004} (open).
The self-interaction corrected PBE-SIC/2 calculation is in close agreement with experimental measurements as well as the high-level calculations, 
while the PBE results are qualitatively incorrect.
Open diamonds show results of calculations where the PBE energy is evaluated for the PBE-SIC/2 electron density, showing that the
dominant error is in the self-interaction rather than in the electron density.
}
\label{fig:fig1}
\end{center}
\end{figure}

Fig.~\ref{fig:fig1} shows the energy of the three lowest energy electronic states of the Mn dimer as a function of the distance between the atoms,
calculated with the PBE functional with and without self-interaction correction.
Three relevant, low lying electronic states are found in the calculations: two ferromagnetic states, ${}^{11}\Sigma^{+}_{u}$ and ${}^{11}\Pi_{u}$, 
and an antiferromagnetic ${}^{1}\Sigma^{+}_g$ state.
The self-interaction corrected functional
predicts the antiferromagnetic state to be the ground state, and gives a
binding energy of 0.18 eV and bond length of 3.32 {\AA}, in close agreement with the experimental results as well as the high-level wave function calculations.
The ferromagnetic ${}^{11}\Sigma^{+}_{u}$ state is only 0.04 eV higher in energy at the optimal bond length.
A vibrational frequency of 69 cm$^{-1}$ is obtained from 
the ground state energy curve, in good agreement with the 
experimental measurements.\cite{bier1988,kirkwood1991}
The magnetic coupling constant is calculated using the relationship\cite{barborini2016}
\begin{equation}
    J = \frac{E_{AF} - E_{F}}{\left<S^{2}_{F}\right> - \left<S^{2}_{AF}\right>},
\end{equation}
where $E_{AF}$ and $E_{F}$ are the energy of the antiferromagnetic and ferromagnetic states, 
and $<S^2>$ is evaluated taking into account spin contamination.\cite{wang1995} 
This gives J= 
-6.0 cm$^{-1}$, 
close to the value obtained in the MCQDPT2 calculations.\cite{yamamoto2006}
The scaling of the self-interaction correction is not important in this case. When the full correction is used as in Eqn.~(5), similar results are obtained, namely
bond energy of 0.12 eV, bond length of 3.5 {\AA} and vibrational frequency of 56 cm$^{-1}$.
Most importantly, the self-interaction correction, scaled or not scaled, gives the right antiferromagnetic ground state. 

Different results are obtained with the uncorrected PBE functional, as can be seen from Fig.~\ref{fig:fig1}.
There, the ferromagnetic state ${}^{11}\Pi_{u}$ is lowest in energy 
and even the other ferromagnetic state, ${}^{11}\Sigma^{+}_{u}$, turns out to be lower than the antiferromagnetic, ${}^{1}\Sigma_g^+$ state at the 
experimental bond length. 
The calculated binding energy in the ground state is 0.90 eV with a bond length of 2.6 {\AA} in strong disagreement with best estimates. 
These results are consistent with previously reported calculations using semi-local functionals\cite{barborini2016}. 
At a large distance between the Mn atoms, the ${}^{11}\Pi_{u}$ state becomes higher in energy than the others as it dissociates into a 
$d^{\,6}s^1$ configuration for one of the Mn atoms.
Calculations with the meta-GGA SCAN functional\cite{sun2015} were also carried out using the VASP software\cite{kresse1996}
and the results are qualitatively similar to the PBE results in that they also give the ferromagnetic ${}^{11}\Pi_{u}$ state as the ground electronic state,
but the binding energy is smaller than for PBE, 0.46 eV.
The SCAN functional can produce an antiferromagnetic ground state if a U-term is added, effectively mimicking a self-interaction correction 
(see supporting information in Ref.\cite{pulkkinen2020}).

In order to analyse this failure of the semi-local functional calculations,
a molecular orbital diagram is shown in Fig. 2.   
Interestingly, the occupation of minority-spin molecular orbitals turns out to play an important role here. 
In the PBE calculation, the bonding minority spin orbital $\pi(3d)$ is lower in energy than the anti-bonding $\sigma^{*}(4s)$, and this leads to $d$-$d$ bond formation
in the minority spin states.
When the self-interaction correction is applied, the relative energy of $\pi(3d)$ and $\sigma^{*}(4s)$ is reversed and the anti-bonding spin-orbital is occupied
instead.
The relative energy of the $d$ and $s$ atomic orbitals is an important issue here. 
\begin{figure}[!t]
\begin{center}
\includegraphics[width = 0.7\columnwidth]{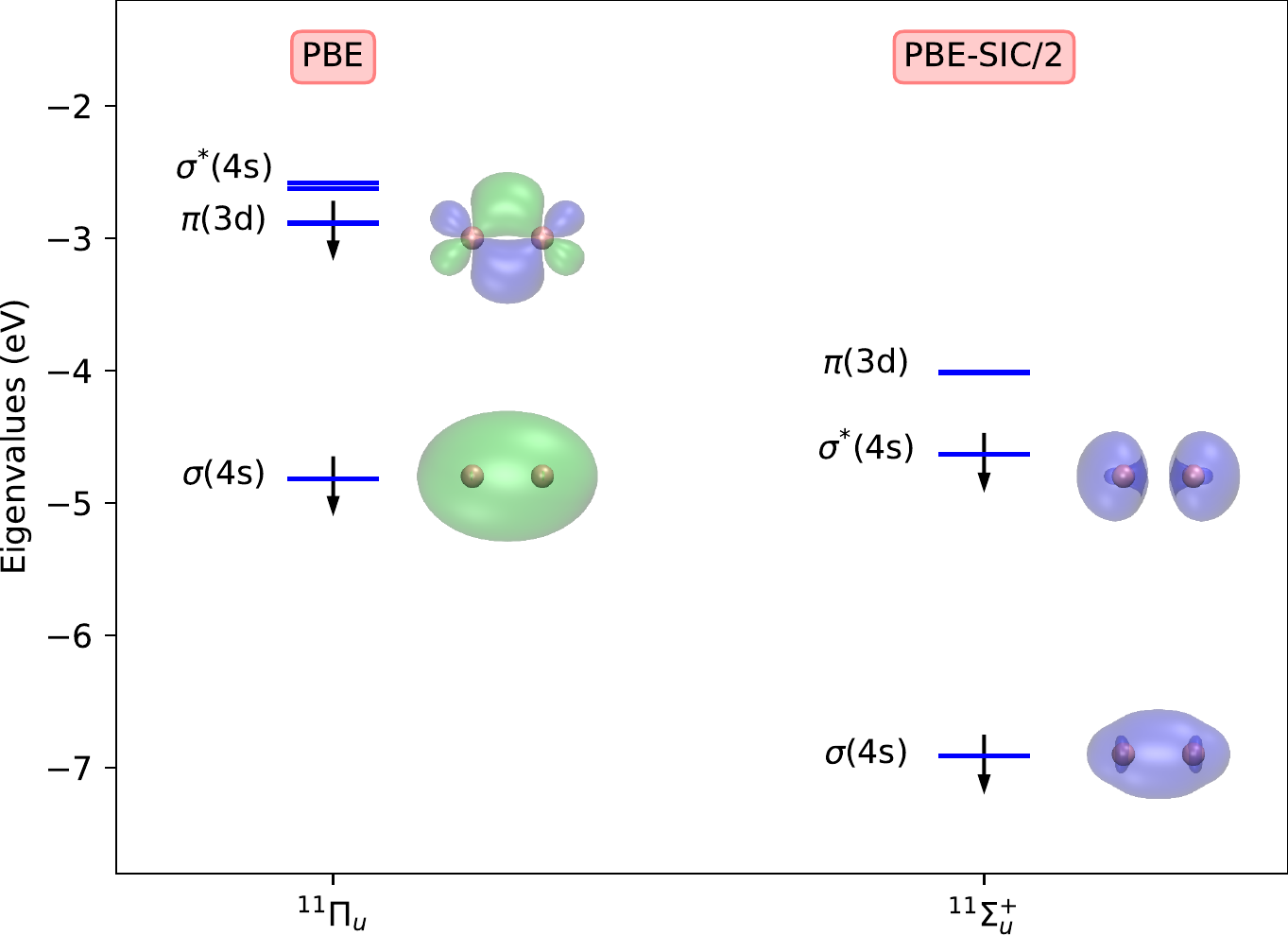}
\caption{Molecular orbital diagram for 
the lowest energy ferromagnetic state of the Mn dimer calculated with the PBE functional and with the self-interaction corrected PBE-SIC/2 functional, 
based on orbital energy of the canonical orbitals.
In the PBE calculations, the bonding $\pi(3d)$ minority spin orbital has lower energy than the antibonding $\sigma^\ast(4s)$ orbital, 
and becomes populated, resulting in a ${}^{11}\Pi_{u}$ ground state for the Mn dimer.
When self-interaction correction is applied, in the PBE-SIC/2 functional, the relative energy of these two molecular spin-orbitals is reversed, and 
$\sigma^\ast(4s)$ becomes populated resulting in a ${}^{11}\Sigma^{+}_{u}$ state of the dimer. 
The magnetic coupling then makes the ${}^{1}\Sigma_g^+$ state slightly lower in energy than the ${}^{11}\Sigma^{+}_{u}$ state, by 0.04 eV,
resulting in an antiferromagnetic ground state in the PBE-SIC/2 calculation.
The surfaces illustrating the molecular spin-orbitals in the insets correspond to a value of $\pm$0.08 {\AA}$^{-3/2}$, where
different colors indicate the sign for the PBE orbitals, but for PBE-SIC/2 the amplitude of the orbitals is shown as they are complex. 
}
\label{fig:fig2}
\end{center}
\end{figure}


%

%

\begin{figure}[!t]
\begin{center}
\includegraphics[width = 0.8\columnwidth]{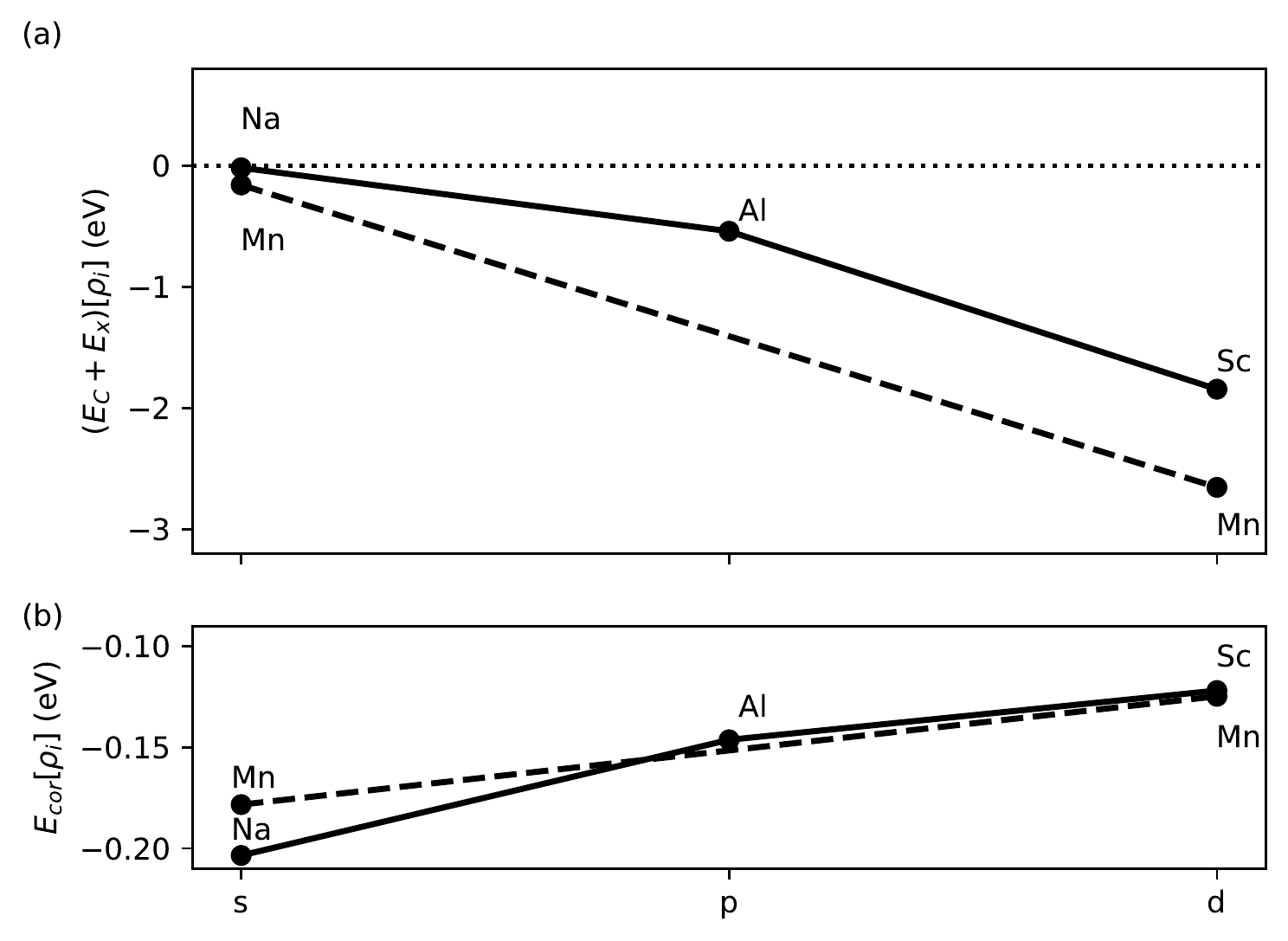}
\caption{
(a)
Net self-interaction, $ E_\mathrm{C}[\rho_i] +  E_{x}[\rho_i, 0]$, evaluated for atomic orbitals using orbital densities obtained from PBE calculations of the atoms.
Results are shown for the Mn atom, as well as a few other atoms (Na, Al, Sc) to illustrate the trend of increasing magnitude of the
negative net self-interaction as the orbital angular momentum increases. 
For $s$ orbitals, the negative self-interaction energy in the exchange part of the PBE functional compensates 
well the positive self-interaction energy in the classical Coulomb part,
but for $d$ orbitals it overcorrects.
The $d$ orbitals in the Mn atom are, therefore, too low in energy compared to 
the $s$ orbitals and this leads to incorrect ordering of the bonding $\pi(3d)$ molecular spin-orbital 
and the anti-bonding $\sigma^\ast(4s)$ spin-orbital in the minority-spin channel, as illustrated in Fig. 2.
(b)
Self-interaction in the correlation part of the PBE functional evaluated for the atomic orbitals as in (a). 
Here, the largest contribution is obtained for the $s$ orbitals, but this contribution is an order of magnitude smaller than the one shown in (a).
}
\label{fig:fig3}
\end{center}
\end{figure}

It is well known that semi-local DFT functionals do not describe well the energy balance between localized and delocalized electrons, and this 
is reflected in the relative energy of $d$ and $s$ atomic orbitals.  
The repulsive self-interaction error in the classical Coulomb interaction is larger the more localized the electrons are.
One might, therefore, expect that $d$ electrons are calculated to be high in energy compared with $s$ electrons. 
However, the results of the calculations presented here using the PBE functional show the opposite trend, as illustrated in 
the molecular orbital diagram of Fig. 2 where the PBE functional ends up placing electrons in a molecular spin-orbital formed 
from $d$ atomic orbitals rather than the one formed from $s$ atomic orbitals. 
The answer lies in the estimate of the self-interaction in the exchange-correlation part of the PBE functional, 
a contribution that should cancel out the repulsive self-interaction in the classical Coulomb term.
This estimate is based on an analysis of smoothly varying electron gas.
The larger the deviation is from the uniform electron gas, the larger an error can be expected in the exchange-correlation functional. 
The extent to which the self-interaction is cancelled out by the two contributions for the various atomic orbitals in the Mn atom as well as a few other atoms
is shown in Fig.~\ref{fig:fig3}.
%
%
The cancellation of the self-interaction energy for the 3$s$ atomic orbital is quite good, the net self-interaction 
being only 0.03 eV. 
But, the net self-interaction has a larger magnitude and is negative for the 3$d$ orbitals of the Mn atom, -2.8 eV.  
This might be related to the fact that the $d$ electrons exhibit a multi-center character which the semi-local functionals often fail to describe.\cite{Shahi2019}
Similar trend is observed for orbitals of other atoms, as shown in the figure. 
The self-interaction correction in the semi-local exchange-correlation functional is, therefore, an overcorrection and 
makes the $d$ atomic orbitals too low in energy as compared with the $s$ atomic orbitals. This leads to the population of a bonding $\pi(3d)$ molecular spin-orbital in the 
Mn dimer instead of an anti-bonding $\sigma^\ast(4s)$ spin-orbital. As a result, the Mn dimer is overbound in an incorrect electronic ground state in the semi-local DFT calculations.

%
%
%
%
%


There are two aspects of the error in DFT calculations:\cite{Kim2013} 
(1) an error in the self-consistent electron density, and 
(2) an error in the energy obtained for a given, possibly correct, electron density. 
A calculation using the PBE functional with the PBE-SIC/2 electron density as input for the Mn dimer is also shown in Fig. 1.
The binding energy for both ferromagnetic states is reduced, but the ${}^{11}\Pi_{u}$ state is still the lowest energy state and 
there is still large overbinding of the dimer. 
This shows that the self-interaction is the main source of error rather than the electron density.

To analyze this further, the electron interaction terms are evaluated using PBE for each of the Mn$_2$ molecular spin-orbitals separately 
using the optimal PBE-SIC/2 orbital densities (see Table~\ref{tbl:tbl1}). The net 
self-interaction energy estimated for the majority-spin ${}^{11}\Pi_{u}$ and ${}^{11}\Sigma^{+}_{u}$ states differs only by $ca.$ 20 meV so 
the relative energy of these states is not affected 
significantly 
by the self-interaction error.
The reason is that these two states have similar type of bonding,
both involve ten molecular orbitals formed from $d$ electrons and two molecular orbitals formed from $s$ electrons.
However, for the minority-spin states, the magnitude of the self-interaction energy is quite different,
being 1.2 eV larger for the ${}^{11}\Pi_{u}$ state than the ${}^{11}\Sigma^{+}_{u}$ state. 
The reason is that the former involves an orbital formed from $d$ electrons, $\pi(3d)$, as well the $\sigma(4s)$ orbital formed from $s$ electrons,
while the latter involves only orbitals formed from $s$ electrons.
As a result, the subtraction of the self-interaction error from the PBE 
functional
has a large effect and reverses the relative energy of the ${}^{11}\Pi_{u}$ and ${}^{11}\Sigma^{+}_{u}$ states.
 
\begin{table}[H]
\caption{\label{tab: sic_spins} The sum of the classical Coulomb and exchange-correlation self-interaction 
energy evaluated with PBE for each molecular spin-orbital separately, 
using optimal spin-orbitals from the PBE-SIC/2 calculation. 
Only the 14 valence electrons are included. 
For the majority-spin orbitals, the difference is just  0.02 eV, 
because they both involve ten molecular orbitals formed from $d$ electrons and two molecular orbitals formed from $s$ electrons. 
But, for the minority-spin, the ${}^{11}\Sigma^{+}_{u}$ state involves two orbitals formed from $s$ electrons, $\sigma(4s)$ and $\sigma^\ast(4s)$, 
while the ${}^{11}\Pi_{u}$ state involves an orbital formed from $d$ electrons, $\pi(3d)$, as well as the $\sigma(4s)$ orbital. 
The self-interaction has, therefore, different magnitude for the minority-spin, being 1.2 eV larger for the ${}^{11}\Pi_{u}$ state 
than for the ${}^{11}\Sigma^{+}_{u}$ and, thereby, reverses the relative stability of these states.
}
\begin{tabular}{cccccccc}
 \hline       
        &\multicolumn{3}{c}{Majority spin}    && \multicolumn{3}{c}{Minority spin} \\
 

        & $\sum E_C[\rho_i]$ & $\sum E_{xc}[\rho_i]$  & total$_{maj}$   &&  $\sum E_C[\rho_i]$  & $\sum E_{xc}[\rho_i]$  & total$_{min}$ \\
\hline
${}^{11}\Pi_{u}$  &114.9 &-126.0  & -11.10         &&         8.8 &-10.6  &-1.8\\
${}^{11}\Sigma^{+}_{u}$  &118.6 & -129.7&  -11.08     &&      7.2 &-7.8& -0.6\\
\hline
\end{tabular}
\label{tbl:tbl1}
\end{table}

In summary,
KS-DFT calculations with the LDA, GGA or meta-GGA functional approximations give qualitatively incorrect results for the Mn dimer.
The reason for this failure is not strong correlation but rather the  
one-electron self-interaction error that results from incomplete cancellation of the non-local self-interaction that is necessarily included in 
an estimate of the classical Coulomb interaction
using only the total electron density as input and the semi-local estimate of the compensating self-interaction in the exchange correlation term in these functionals. 
It turns out that the net self-interaction error is particularly large for the $d$-electrons and artificially lowers their energy with respect to that of $s$-electrons
because of an overestimate of the self-interaction in the 
semi-local
exchange correlation functionals.  As a result, an electronic state that dissociates into an
Mn atom with a $d^{\,6}s^1$ electron configuration becomes the ground state and much too strong bonding is obtained 
because of
constructive
overlap of the $d$-orbitals in the minority spin channel.  
The results presented here show that an explicit self-interaction correction for each spin-orbital applied to the PBE functional accurately describes the 
Mn dimer, predicting bond distance, binding energy, vibrational frequency and magnetic coupling constant 
in close agreement with experimental measurements and high-level quantum chemistry calculations. Since the computational effort in the 
self-interaction corrected calculations scales with system size in the same way as DFT calculations with semi-local functionals, i.e. as the system size to the third power, 
the method can be applied to large systems including extended solids.\cite{Jonsson2011,gudmundsdottir2015}
There are several other formulations of self-interaction correction and different implementations\cite{Perdew2015}
and it would be interesting in future work to see how well they perform in calculations of the Mn dimer.

\begin{acknowledgement}
This work was supported by the Icelandic Research Fund and the Academy of Finland. 
AVI is supported by a doctoral fellowship from the University of Iceland and thanks Tuomas Rossi and Valery Uzdin for helpful discussions.
\end{acknowledgement}




\bibliography{re-revisedmanuscript-Mn2}

\providecommand{\latin}[1]{#1}
\makeatletter
\providecommand{\doi}
  {\begingroup\let\do\@makeother\dospecials
  \catcode`\{=1 \catcode`\}=2 \doi@aux}
\providecommand{\doi@aux}[1]{\endgroup\texttt{#1}}
\makeatother
\providecommand*\mcitethebibliography{\thebibliography}
\csname @ifundefined\endcsname{endmcitethebibliography}
  {\let\endmcitethebibliography\endthebibliography}{}
\begin{mcitethebibliography}{48}
\providecommand*\natexlab[1]{#1}
\providecommand*\mciteSetBstSublistMode[1]{}
\providecommand*\mciteSetBstMaxWidthForm[2]{}
\providecommand*\mciteBstWouldAddEndPuncttrue
  {\def\EndOfBibitem{\unskip.}}
\providecommand*\mciteBstWouldAddEndPunctfalse
  {\let\EndOfBibitem\relax}
\providecommand*\mciteSetBstMidEndSepPunct[3]{}
\providecommand*\mciteSetBstSublistLabelBeginEnd[3]{}
\providecommand*\EndOfBibitem{}
\mciteSetBstSublistMode{f}
\mciteSetBstMaxWidthForm{subitem}{(\alph{mcitesubitemcount})}
\mciteSetBstSublistLabelBeginEnd
  {\mcitemaxwidthsubitemform\space}
  {\relax}
  {\relax}

\bibitem[Hanson and Berliner(2010)Hanson, and Berliner]{Book_EPR_bio}
Hanson,~G., Berliner,~L., Eds. \emph{Metals in Biology: Applications of
  High-Resolution EPR to Metalloenzymes}; Springer: New York, NY, 2010;
  Vol.~29\relax
\mciteBstWouldAddEndPuncttrue
\mciteSetBstMidEndSepPunct{\mcitedefaultmidpunct}
{\mcitedefaultendpunct}{\mcitedefaultseppunct}\relax
\EndOfBibitem
\bibitem[Dismukes(1996)]{Dismukes_chem_rev}
Dismukes,~G.~C. Manganese Enzymes with Binuclear Active Sites. \emph{Chem.
  Rev.} \textbf{1996}, \emph{96}, 2909--2926\relax
\mciteBstWouldAddEndPuncttrue
\mciteSetBstMidEndSepPunct{\mcitedefaultmidpunct}
{\mcitedefaultendpunct}{\mcitedefaultseppunct}\relax
\EndOfBibitem
\bibitem[Rivoal \latin{et~al.}(1982)Rivoal, Emampour, Zeringue, and
  Vala]{rivoal1982}
Rivoal,~J.-C.; Emampour,~J.~S.; Zeringue,~K.~J.; Vala,~M. Ground-state exchange
  energy of the Mn$_2$ antiferromagnetic molecule. \emph{Chem. Phys. Lett.}
  \textbf{1982}, \emph{92}, 313--316\relax
\mciteBstWouldAddEndPuncttrue
\mciteSetBstMidEndSepPunct{\mcitedefaultmidpunct}
{\mcitedefaultendpunct}{\mcitedefaultseppunct}\relax
\EndOfBibitem
\bibitem[Baumann \latin{et~al.}(1983)Baumann, Zee, Bhat, and Jr.]{baumann1983}
Baumann,~C.~A.; Zee,~R. J.~V.; Bhat,~S.~V.; Jr.,~W.~W. ESR of Mn$_2$ and Mn$_5$
  molecules in rare-gas matrices. \emph{J. Chem. Phys.} \textbf{1983},
  \emph{78}, 190--199\relax
\mciteBstWouldAddEndPuncttrue
\mciteSetBstMidEndSepPunct{\mcitedefaultmidpunct}
{\mcitedefaultendpunct}{\mcitedefaultseppunct}\relax
\EndOfBibitem
\bibitem[Kirkwood \latin{et~al.}(1991)Kirkwood, Bier, Thompson, Haslett, Huber,
  and Moskovits]{kirkwood1991}
Kirkwood,~A.~D.; Bier,~K.~D.; Thompson,~J.~K.; Haslett,~T.~L.; Huber,~A.~S.;
  Moskovits,~M. Ultraviolet-visible and Raman spectroscopy of diatomic
  manganese isolated in rare-gas matrixes. \emph{J. Phys. Chem.} \textbf{1991},
  \emph{95}, 2644--2652\relax
\mciteBstWouldAddEndPuncttrue
\mciteSetBstMidEndSepPunct{\mcitedefaultmidpunct}
{\mcitedefaultendpunct}{\mcitedefaultseppunct}\relax
\EndOfBibitem
\bibitem[Kant \latin{et~al.}(1968)Kant, Lin, and Strauss]{kant1968}
Kant,~A.; Lin,~S.; Strauss,~B. Dissociation Energy of Mn$_2$. \emph{J. Chem.
  Phys.} \textbf{1968}, \emph{49}, 1983--1985\relax
\mciteBstWouldAddEndPuncttrue
\mciteSetBstMidEndSepPunct{\mcitedefaultmidpunct}
{\mcitedefaultendpunct}{\mcitedefaultseppunct}\relax
\EndOfBibitem
\bibitem[Cheeseman \latin{et~al.}(1990)Cheeseman, Van~Zee, Flanagan, and
  Weltner]{cheeseman1990}
Cheeseman,~M.; Van~Zee,~R.~J.; Flanagan,~H.~L.; Weltner,~W. Transition metal
  diatomics: Mn$_2$, Mn$^+_2$, CrMn. \emph{J. Chem. Phys.} \textbf{1990},
  \emph{92}, 1553--1559\relax
\mciteBstWouldAddEndPuncttrue
\mciteSetBstMidEndSepPunct{\mcitedefaultmidpunct}
{\mcitedefaultendpunct}{\mcitedefaultseppunct}\relax
\EndOfBibitem
\bibitem[Bier \latin{et~al.}(1988)Bier, Haslett, Kirkwood, and
  Moskovits]{bier1988}
Bier,~K.~D.; Haslett,~T.~L.; Kirkwood,~A.~D.; Moskovits,~M. The resonance Raman
  and visible absorbance spectra of matrix isolated Mn$_2$ and Mn$_3$. \emph{J.
  Chem. Phys.} \textbf{1988}, \emph{89}, 6--12\relax
\mciteBstWouldAddEndPuncttrue
\mciteSetBstMidEndSepPunct{\mcitedefaultmidpunct}
{\mcitedefaultendpunct}{\mcitedefaultseppunct}\relax
\EndOfBibitem
\bibitem[Yamamoto \latin{et~al.}(2006)Yamamoto, Tatewaki, Moriyama, and
  Nakano]{yamamoto2006}
Yamamoto,~S.; Tatewaki,~H.; Moriyama,~H.; Nakano,~H. A study of the ground
  state of manganese dimer using quasidegenerate perturbation theory. \emph{J.
  Chem. Phys.} \textbf{2006}, \emph{124}, 124302\relax
\mciteBstWouldAddEndPuncttrue
\mciteSetBstMidEndSepPunct{\mcitedefaultmidpunct}
{\mcitedefaultendpunct}{\mcitedefaultseppunct}\relax
\EndOfBibitem
\bibitem[Wang and Chen(2004)Wang, and Chen]{wang2004}
Wang,~B.; Chen,~Z. Magnetic coupling interaction under different spin
  multiplets in neutral manganese dimer: CASPT2 theoretical investigation.
  \emph{Chem. Phys. Lett.} \textbf{2004}, \emph{387}, 395 -- 399\relax
\mciteBstWouldAddEndPuncttrue
\mciteSetBstMidEndSepPunct{\mcitedefaultmidpunct}
{\mcitedefaultendpunct}{\mcitedefaultseppunct}\relax
\EndOfBibitem
\bibitem[Camacho \latin{et~al.}(2008)Camacho, Yamamoto, and Witek]{camacho2008}
Camacho,~C.; Yamamoto,~S.; Witek,~H.~A. Choosing a proper complete active space
  in calculations for transition metal dimers: ground state of Mn$_2$
  revisited. \emph{Phys. Chem. Chem. Phys.} \textbf{2008}, \emph{10},
  5128--5134\relax
\mciteBstWouldAddEndPuncttrue
\mciteSetBstMidEndSepPunct{\mcitedefaultmidpunct}
{\mcitedefaultendpunct}{\mcitedefaultseppunct}\relax
\EndOfBibitem
\bibitem[Angeli \latin{et~al.}(2008)Angeli, Cavallini, and
  Cimiraglia]{angeli2008}
Angeli,~C.; Cavallini,~A.; Cimiraglia,~R. An ab initio multireference
  perturbation theory study on the manganese dimer. \emph{J. Chem. Phys.}
  \textbf{2008}, \emph{128}, 244317\relax
\mciteBstWouldAddEndPuncttrue
\mciteSetBstMidEndSepPunct{\mcitedefaultmidpunct}
{\mcitedefaultendpunct}{\mcitedefaultseppunct}\relax
\EndOfBibitem
\bibitem[Buchachenko \latin{et~al.}(2010)Buchachenko, Cha\l{}asi\'nski, and
  Szcz\c{e}\'niak]{buchachenko2010}
Buchachenko,~A.~A.; Cha\l{}asi\'nski,~G.; Szcz\c{e}\'niak,~M.~M. Electronic
  structure and spin coupling of the manganese dimer: The state of the art of
  ab initio approach. \emph{J. Chem. Phys.} \textbf{2010}, \emph{132},
  024312\relax
\mciteBstWouldAddEndPuncttrue
\mciteSetBstMidEndSepPunct{\mcitedefaultmidpunct}
{\mcitedefaultendpunct}{\mcitedefaultseppunct}\relax
\EndOfBibitem
\bibitem[Kohn and Sham(1965)Kohn, and Sham]{kohn1965}
Kohn,~W.; Sham,~L.~J. Self-Consistent Equations Including Exchange and
  Correlation Effects. \emph{Phys. Rev.} \textbf{1965}, \emph{140},
  A1133--A1138\relax
\mciteBstWouldAddEndPuncttrue
\mciteSetBstMidEndSepPunct{\mcitedefaultmidpunct}
{\mcitedefaultendpunct}{\mcitedefaultseppunct}\relax
\EndOfBibitem
\bibitem[Pederson \latin{et~al.}(1998)Pederson, Reuse, and
  Khanna]{pederson1998}
Pederson,~M.~R.; Reuse,~F.; Khanna,~S.~N. Magnetic transition in
  ${\mathrm{Mn}}_{n}$ $(n=2-8)$ clusters. \emph{Phys. Rev. B} \textbf{1998},
  \emph{58}, 5632--5636\relax
\mciteBstWouldAddEndPuncttrue
\mciteSetBstMidEndSepPunct{\mcitedefaultmidpunct}
{\mcitedefaultendpunct}{\mcitedefaultseppunct}\relax
\EndOfBibitem
\bibitem[Yanagisawa \latin{et~al.}(2000)Yanagisawa, Tsuneda, and
  Hirao]{Yanagisawa2000}
Yanagisawa,~S.; Tsuneda,~T.; Hirao,~K. {An investigation of density
  functionals: The first-row transition metal dimer calculations}. \emph{J.
  Chem. Phys.} \textbf{2000}, \emph{112}, 545--553\relax
\mciteBstWouldAddEndPuncttrue
\mciteSetBstMidEndSepPunct{\mcitedefaultmidpunct}
{\mcitedefaultendpunct}{\mcitedefaultseppunct}\relax
\EndOfBibitem
\bibitem[Barborini(2016)]{barborini2016}
Barborini,~M. Neutral, Anionic, and Cationic Manganese Dimers through Density
  Functional Theory. \emph{J. Phys. Chem. A} \textbf{2016}, \emph{120},
  1716--1726\relax
\mciteBstWouldAddEndPuncttrue
\mciteSetBstMidEndSepPunct{\mcitedefaultmidpunct}
{\mcitedefaultendpunct}{\mcitedefaultseppunct}\relax
\EndOfBibitem
\bibitem[Sun \latin{et~al.}(2015)Sun, Ruzsinszky, and Perdew]{sun2015}
Sun,~J.; Ruzsinszky,~A.; Perdew,~J.~P. Strongly Constrained and Appropriately
  Normed Semilocal Density Functional. \emph{Phys. Rev. Lett.} \textbf{2015},
  \emph{115}, 036402\relax
\mciteBstWouldAddEndPuncttrue
\mciteSetBstMidEndSepPunct{\mcitedefaultmidpunct}
{\mcitedefaultendpunct}{\mcitedefaultseppunct}\relax
\EndOfBibitem
\bibitem[Pantazis(1999)]{Pantazis_Inorganics}
Pantazis,~D. Assessment of Double-Hybrid Density Functional Theory for Magnetic
  Exchange Coupling in Manganese Complexes. \emph{Inorganics} \textbf{1999},
  \emph{7}, 57\relax
\mciteBstWouldAddEndPuncttrue
\mciteSetBstMidEndSepPunct{\mcitedefaultmidpunct}
{\mcitedefaultendpunct}{\mcitedefaultseppunct}\relax
\EndOfBibitem
\bibitem[Perdew \latin{et~al.}(2021)Perdew, Ruzsinszky, Sun, Nepal, and
  Kaplan]{perdew2021}
Perdew,~J.~P.; Ruzsinszky,~A.; Sun,~J.; Nepal,~N.~K.; Kaplan,~A.~D.
  {Interpretations of ground-state symmetry breaking and strong correlation in
  wavefunction and density functional theories}. \emph{Proc. Natl. Acad. Sci.
  USA} \textbf{2021}, \emph{118}, 1--6\relax
\mciteBstWouldAddEndPuncttrue
\mciteSetBstMidEndSepPunct{\mcitedefaultmidpunct}
{\mcitedefaultendpunct}{\mcitedefaultseppunct}\relax
\EndOfBibitem
\bibitem[Yamanaka \latin{et~al.}(2007)Yamanaka, Ukai, Nakata, Takeda, Shoji,
  Kawakami, Takada, and Yamaguchi]{yamanaka2007}
Yamanaka,~S.; Ukai,~T.; Nakata,~K.; Takeda,~R.; Shoji,~M.; Kawakami,~T.;
  Takada,~T.; Yamaguchi,~K. Density functional study of manganese dimer.
  \emph{Int. J. Quantum Chem.} \textbf{2007}, \emph{107}, 3178--3190\relax
\mciteBstWouldAddEndPuncttrue
\mciteSetBstMidEndSepPunct{\mcitedefaultmidpunct}
{\mcitedefaultendpunct}{\mcitedefaultseppunct}\relax
\EndOfBibitem
\bibitem[Perdew and Zunger(1981)Perdew, and Zunger]{perdew1981}
Perdew,~J.~P.; Zunger,~A. Self-interaction correction to density-functional
  approximations for many-electron systems. \emph{Phys. Rev. B} \textbf{1981},
  \emph{23}, 5048--5079\relax
\mciteBstWouldAddEndPuncttrue
\mciteSetBstMidEndSepPunct{\mcitedefaultmidpunct}
{\mcitedefaultendpunct}{\mcitedefaultseppunct}\relax
\EndOfBibitem
\bibitem[Kl{\"u}pfel \latin{et~al.}(2012)Kl{\"u}pfel, Kl{\"u}pfel, Tsemekhman,
  and J{\'o}nsson]{klupfel2012b}
Kl{\"u}pfel,~P.; Kl{\"u}pfel,~S.; Tsemekhman,~K.; J{\'o}nsson,~H. Optimization
  of functionals of orthonormal functions in the absence of unitary invariance.
  \emph{Lecture Notes in Computer Science} \textbf{2012}, \emph{7134}, 23\relax
\mciteBstWouldAddEndPuncttrue
\mciteSetBstMidEndSepPunct{\mcitedefaultmidpunct}
{\mcitedefaultendpunct}{\mcitedefaultseppunct}\relax
\EndOfBibitem
\bibitem[J\'onsson(2011)]{Jonsson2011}
J\'onsson,~H. {Simulation of surface processes}. \emph{Proc. Natl. Acad. Sci.}
  \textbf{2011}, \emph{108}, 944--949\relax
\mciteBstWouldAddEndPuncttrue
\mciteSetBstMidEndSepPunct{\mcitedefaultmidpunct}
{\mcitedefaultendpunct}{\mcitedefaultseppunct}\relax
\EndOfBibitem
\bibitem[Kl{\"u}pfel \latin{et~al.}(2012)Kl{\"u}pfel, Kl{\"u}pfel, and
  J{\'o}nsson]{klupfel2012}
Kl{\"u}pfel,~S.; Kl{\"u}pfel,~P.; J{\'o}nsson,~H. The effect of the
  Perdew-Zunger self-interaction correction to density functionals on the
  energetics of small molecules. \emph{J. Chem. Phys.} \textbf{2012},
  \emph{137}, 124102\relax
\mciteBstWouldAddEndPuncttrue
\mciteSetBstMidEndSepPunct{\mcitedefaultmidpunct}
{\mcitedefaultendpunct}{\mcitedefaultseppunct}\relax
\EndOfBibitem
\bibitem[Gudmundsd{\'o}ttir \latin{et~al.}(2015)Gudmundsd{\'o}ttir,
  J{\'o}nsson, and J{\'o}nsson]{gudmundsdottir2015}
Gudmundsd{\'o}ttir,~H.; J{\'o}nsson,~E.~O.; J{\'o}nsson,~H. Calculations of Al
  dopant in $\alpha$-quartz using a variational implementation of the
  Perdew-Zunger self-interaction correction. \emph{New J. Phys.} \textbf{2015},
  \emph{17}, 083006\relax
\mciteBstWouldAddEndPuncttrue
\mciteSetBstMidEndSepPunct{\mcitedefaultmidpunct}
{\mcitedefaultendpunct}{\mcitedefaultseppunct}\relax
\EndOfBibitem
\bibitem[Perdew \latin{et~al.}(1996)Perdew, Burke, and Ernzerhof]{perdew1996}
Perdew,~J.~P.; Burke,~K.; Ernzerhof,~M. Generalized Gradient Approximation Made
  Simple. \emph{Phys. Rev. Lett.} \textbf{1996}, \emph{77}, 3865--3868\relax
\mciteBstWouldAddEndPuncttrue
\mciteSetBstMidEndSepPunct{\mcitedefaultmidpunct}
{\mcitedefaultendpunct}{\mcitedefaultseppunct}\relax
\EndOfBibitem
\bibitem[Kl\"upfel \latin{et~al.}(2011)Kl\"upfel, Kl\"upfel, and
  J\'onsson]{klupfel2011}
Kl\"upfel,~S.; Kl\"upfel,~P.; J\'onsson,~H. Importance of complex orbitals in
  calculating the self-interaction-corrected ground state of atoms. \emph{Phys.
  Rev. A} \textbf{2011}, \emph{84}, 050501\relax
\mciteBstWouldAddEndPuncttrue
\mciteSetBstMidEndSepPunct{\mcitedefaultmidpunct}
{\mcitedefaultendpunct}{\mcitedefaultseppunct}\relax
\EndOfBibitem
\bibitem[Lehtola and J\'onsson(2013)Lehtola, and J\'onsson]{Lehtola2013}
Lehtola,~S.; J\'onsson,~H. {Unitary optimization of localized molecular
  orbitals}. \emph{J. Chem. Theo. Comput.} \textbf{2013}, \emph{9},
  5365--5372\relax
\mciteBstWouldAddEndPuncttrue
\mciteSetBstMidEndSepPunct{\mcitedefaultmidpunct}
{\mcitedefaultendpunct}{\mcitedefaultseppunct}\relax
\EndOfBibitem
\bibitem[Lehtola \latin{et~al.}(2016)Lehtola, Head-Gordon, and
  J\'onsson]{Lehtola2016}
Lehtola,~S.; Head-Gordon,~M.; J\'onsson,~H. {Complex orbitals, multiple local
  minima, and symmetry breaking in Perdew–Zunger self-interaction corrected
  density functional theory calculations}. \emph{J. Chem. Theo. Comput.}
  \textbf{2016}, \emph{12}, 3195–3207\relax
\mciteBstWouldAddEndPuncttrue
\mciteSetBstMidEndSepPunct{\mcitedefaultmidpunct}
{\mcitedefaultendpunct}{\mcitedefaultseppunct}\relax
\EndOfBibitem
\bibitem[Enkovaara and {\it et. al.}(2010)Enkovaara, and {\it et.
  al.}]{enkovaara2010}
Enkovaara,~J.; {\it et. al.}, Electronic structure calculations with GPAW: a
  real-space implementation of the projector augmented-wave method. \emph{J.
  Phys. Condens. Matter} \textbf{2010}, \emph{22}, 253202\relax
\mciteBstWouldAddEndPuncttrue
\mciteSetBstMidEndSepPunct{\mcitedefaultmidpunct}
{\mcitedefaultendpunct}{\mcitedefaultseppunct}\relax
\EndOfBibitem
\bibitem[Bl\"ochl(1994)]{blochl1994}
Bl\"ochl,~P.~E. Projector augmented-wave method. \emph{Phys. Rev. B}
  \textbf{1994}, \emph{50}, 17953--17979\relax
\mciteBstWouldAddEndPuncttrue
\mciteSetBstMidEndSepPunct{\mcitedefaultmidpunct}
{\mcitedefaultendpunct}{\mcitedefaultseppunct}\relax
\EndOfBibitem
\bibitem[Weigend and Ahlrichs(2005)Weigend, and Ahlrichs]{Weigend2005}
Weigend,~F.; Ahlrichs,~R. {Balanced basis sets of split valence{,} triple zeta
  valence and quadruple zeta valence quality for H to Rn: Design and assessment
  of accuracy}. \emph{Phys. Chem. Chem. Phys.} \textbf{2005}, \emph{7},
  3297--3305\relax
\mciteBstWouldAddEndPuncttrue
\mciteSetBstMidEndSepPunct{\mcitedefaultmidpunct}
{\mcitedefaultendpunct}{\mcitedefaultseppunct}\relax
\EndOfBibitem
\bibitem[Rappoport and Furche(2010)Rappoport, and Furche]{Rappoport2010}
Rappoport,~D.; Furche,~F. {Property-optimized Gaussian basis sets for molecular
  response calculations}. \emph{J. Chem. Phys.} \textbf{2010}, \emph{133},
  134105\relax
\mciteBstWouldAddEndPuncttrue
\mciteSetBstMidEndSepPunct{\mcitedefaultmidpunct}
{\mcitedefaultendpunct}{\mcitedefaultseppunct}\relax
\EndOfBibitem
\bibitem[Feller(1996)]{Feller1996}
Feller,~D. {The role of databases in support of computational chemistry
  calculations}. \emph{J. Comput. Chem.} \textbf{1996}, \emph{17},
  1571--1586\relax
\mciteBstWouldAddEndPuncttrue
\mciteSetBstMidEndSepPunct{\mcitedefaultmidpunct}
{\mcitedefaultendpunct}{\mcitedefaultseppunct}\relax
\EndOfBibitem
\bibitem[Schuchardt \latin{et~al.}(2007)Schuchardt, Didier, Elsethagen, Sun,
  Gurumoorthi, Chase, Li, and Windus]{Schuchardt2007}
Schuchardt,~K.~L.; Didier,~B.~T.; Elsethagen,~T.; Sun,~L.; Gurumoorthi,~V.;
  Chase,~J.; Li,~J.; Windus,~T.~L. {Basis Set Exchange: A Community Database
  for Computational Sciences}. \emph{J. Chem. Inf. Model.} \textbf{2007},
  \emph{47}, 1045--1052\relax
\mciteBstWouldAddEndPuncttrue
\mciteSetBstMidEndSepPunct{\mcitedefaultmidpunct}
{\mcitedefaultendpunct}{\mcitedefaultseppunct}\relax
\EndOfBibitem
\bibitem[Pritchard \latin{et~al.}(2019)Pritchard, Altarawy, Didier, Gibson, and
  Windus]{Pritchard2019}
Pritchard,~B.~P.; Altarawy,~D.; Didier,~B.; Gibson,~T.~D.; Windus,~T.~L. {New
  Basis Set Exchange: An Open, Up-to-Date Resource for the Molecular Sciences
  Community}. \emph{J. Chem. Inf. Model.} \textbf{2019}, \emph{59},
  4814--4820\relax
\mciteBstWouldAddEndPuncttrue
\mciteSetBstMidEndSepPunct{\mcitedefaultmidpunct}
{\mcitedefaultendpunct}{\mcitedefaultseppunct}\relax
\EndOfBibitem
\bibitem[Rossi \latin{et~al.}(2015)Rossi, Lehtola, Sakko, Puska, and
  Nieminen]{rossi2015}
Rossi,~T.; Lehtola,~S.; Sakko,~A.; Puska,~M.; Nieminen,~R. \emph{J. Chem.
  Phys.} \textbf{2015}, \emph{142}, 094114\relax
\mciteBstWouldAddEndPuncttrue
\mciteSetBstMidEndSepPunct{\mcitedefaultmidpunct}
{\mcitedefaultendpunct}{\mcitedefaultseppunct}\relax
\EndOfBibitem
\bibitem[J{\'{o}}nsson \latin{et~al.}(2017)J{\'{o}}nsson, Lehtola, Puska, and
  J{\'{o}}nsson]{Jonsson2017}
J{\'{o}}nsson,~E.~O.; Lehtola,~S.; Puska,~M.; J{\'{o}}nsson,~H. {Theory and
  Applications of Generalized Pipek-Mezey Wannier Functions}. \emph{J. Chem.
  Theo. Comput.} \textbf{2017}, \emph{13}, 460--474\relax
\mciteBstWouldAddEndPuncttrue
\mciteSetBstMidEndSepPunct{\mcitedefaultmidpunct}
{\mcitedefaultendpunct}{\mcitedefaultseppunct}\relax
\EndOfBibitem
\bibitem[Ivanov \latin{et~al.}(2021)Ivanov, J\'onsson\, Vegge, and
  J\'onsson]{DFTmin2021}
Ivanov,~A.~V.; J\'onsson\,~E.; Vegge,~T.; J\'onsson,~H. Implementation of a
  Direct Minimisation Method using Exponential Transformation in the Localised
  Basis Set Approach. \emph{arXiv:2101.12597} \textbf{2021}, \relax
\mciteBstWouldAddEndPunctfalse
\mciteSetBstMidEndSepPunct{\mcitedefaultmidpunct}
{}{\mcitedefaultseppunct}\relax
\EndOfBibitem
\bibitem[Levi \latin{et~al.}(2020)Levi, Ivanov, and J\'onsson]{DFTmin2020}
Levi,~G.; Ivanov,~A.~V.; J\'onsson,~H. {Variational density functional
  calculations of excited states via direct optimization}. \emph{J. Chem. Theo.
  Comput.} \textbf{2020}, \emph{16}, 6968\relax
\mciteBstWouldAddEndPuncttrue
\mciteSetBstMidEndSepPunct{\mcitedefaultmidpunct}
{\mcitedefaultendpunct}{\mcitedefaultseppunct}\relax
\EndOfBibitem
\bibitem[Wang \latin{et~al.}(1995)Wang, Becke, and Smith~Jr.]{wang1995}
Wang,~J.; Becke,~A.~D.; Smith~Jr.,~V.~H. Evaluation of $\left<S^{2}\right>$ in
  restricted, unrestricted Hartree–Fock, and density functional based
  theories. \emph{J. Chem. Phys.} \textbf{1995}, \emph{102}, 3477--3480\relax
\mciteBstWouldAddEndPuncttrue
\mciteSetBstMidEndSepPunct{\mcitedefaultmidpunct}
{\mcitedefaultendpunct}{\mcitedefaultseppunct}\relax
\EndOfBibitem
\bibitem[Kresse and Furthm\"uller(1996)Kresse, and Furthm\"uller]{kresse1996}
Kresse,~G.; Furthm\"uller,~J. Efficient iterative schemes for ab initio
  total-energy calculations using a plane-wave basis set. \emph{Phys. Rev. B}
  \textbf{1996}, \emph{54}, 11169--11186\relax
\mciteBstWouldAddEndPuncttrue
\mciteSetBstMidEndSepPunct{\mcitedefaultmidpunct}
{\mcitedefaultendpunct}{\mcitedefaultseppunct}\relax
\EndOfBibitem
\bibitem[Pulkkinen \latin{et~al.}(2020)Pulkkinen, Barbiellini, Nokelainen,
  Sokolovskiy, Baigutlin, Miroshkina, Zagrebin, Buchelnikov, Lane, Markiewicz,
  Bansil, Sun, Pussi, and L{\"{a}}hderanta]{pulkkinen2020}
Pulkkinen,~A.; Barbiellini,~B.; Nokelainen,~J.; Sokolovskiy,~V.; Baigutlin,~D.;
  Miroshkina,~O.; Zagrebin,~M.; Buchelnikov,~V.; Lane,~C.; Markiewicz,~R.~S.;
  Bansil,~A.; Sun,~J.; Pussi,~K.; L{\"{a}}hderanta,~E. {Coulomb correlation in
  noncollinear antiferromagnetic $\alpha$-Mn}. \emph{Phys. Rev. B}
  \textbf{2020}, \emph{101}, 075115\relax
\mciteBstWouldAddEndPuncttrue
\mciteSetBstMidEndSepPunct{\mcitedefaultmidpunct}
{\mcitedefaultendpunct}{\mcitedefaultseppunct}\relax
\EndOfBibitem
\bibitem[Shahi \latin{et~al.}(2019)Shahi, Bhattarai, Wagle, Santra, Schwalbe,
  Hahn, Kortus, Jackson, Peralta, Trepte, Lehtola, Nepal, Myneni, Neupane,
  Adhikari, Ruzsinszky, Yamamoto, Baruah, Zope, and Perdew]{Shahi2019}
Shahi,~C. \latin{et~al.}  {Stretched or noded orbital densities and
  self-interaction correction in density functional theory}. \emph{J. Chem.
  Phys.} \textbf{2019}, \emph{150}\relax
\mciteBstWouldAddEndPuncttrue
\mciteSetBstMidEndSepPunct{\mcitedefaultmidpunct}
{\mcitedefaultendpunct}{\mcitedefaultseppunct}\relax
\EndOfBibitem
\bibitem[Kim \latin{et~al.}(2013)Kim, Sim, and Burke]{Kim2013}
Kim,~M.-C.; Sim,~E.; Burke,~K. {Understanding and Reducing Errors in Density
  Functional Calculations}. \emph{Phys. Rev. Lett.} \textbf{2013}, \emph{111},
  73003\relax
\mciteBstWouldAddEndPuncttrue
\mciteSetBstMidEndSepPunct{\mcitedefaultmidpunct}
{\mcitedefaultendpunct}{\mcitedefaultseppunct}\relax
\EndOfBibitem
\bibitem[Perdew \latin{et~al.}(2015)Perdew, Ruzsinszky, Sun, and
  Pederson]{Perdew2015}
Perdew,~J.~P.; Ruzsinszky,~A.; Sun,~J.; Pederson,~M.~R. In \emph{Chapter One -
  Paradox of Self-Interaction Correction: How Can Anything So Right Be So
  Wrong?}; Arimondo,~E., Lin,~C.~C., Yelin,~S.~F., Eds.; Advances In Atomic,
  Molecular, and Optical Physics; Academic Press, 2015; Vol.~64; pp 1--14\relax
\mciteBstWouldAddEndPuncttrue
\mciteSetBstMidEndSepPunct{\mcitedefaultmidpunct}
{\mcitedefaultendpunct}{\mcitedefaultseppunct}\relax
\EndOfBibitem
\end{mcitethebibliography}

\end{document}